\documentstyle[aps,epsfig]{revtex}

\begin{document}

\draft
\twocolumn[\hsize\textwidth\columnwidth\hsize\csname@twocolumnfalse\endcsname

\title{Thermodynamics of quantum Heisenberg spin chains }

\author{Tao Xiang}

\address{Research Centre in
Superconductivity, University of Cambridge, Madingley Road,
Cambridge CH3 0HE, United Kingdom}

\address{and Institute of Theoretical Physics, Academia Sinica, 
P.O.Box 2735, Beijing 100080, China}

\date{\today}

\maketitle

\begin{abstract}

Thermodynamic properties of the quantum Heisenberg spin
chains with $S = 1/2$, 1, and 3/2 are investigated using the
transfer-matrix renormalization-group method.  The
temperature dependence of the magnetization, 
susceptibility, specific heat, spin-spin 
correlation length, and several other physical
quantities in a zero or finite applied field are calculated
and compared.  Our data agree well with the Bethe ansatz,
exact diagonalization, and quantum Monte Carlo results and
provide further insight into the quantum effects in the
antiferromagnetic Heisenberg spin chains.

\end{abstract}


]

\section{Introduction}

There are a lot of quasi-one dimensional compounds 
whose behaviors can be adequately described within
the framework of interacting spin chains governed by the
Heisenberg model.  Extensive studies on this model have shed
light on the quantum nature of spin dynamics.  In 1983,
Haldane predicted that the one-dimensional Heisenberg
antiferromagnetic model with integer spin has an excitation
gap and a finite correlation length\cite{Haldane83}.  Since then
a great amount of experimental and theoretical effort has been
made towards understanding the difference between half-integer
and integer spin chains.

In this paper we report results of a transfer-matrix
renormalization-group (TMRG)
\cite{Nishino95,Bursill96,Wang97} study for the
thermodynamics of quantum Heisenberg spin chains.  We have
calculated the magnetic susceptibility, specific heat,
spin-spin correlation length, and other experimentally
relevant quantities as functions of temperature and applied
magnetic field for the S=1/2, 1, and 3/2 spin chains.

The Heisenberg model is defined by the Hamiltonian
\begin{equation} 
{\hat H} = \sum_{i}^{N}h_i ;\quad 
h_i  = J {\bf S}_i \cdot {\bf S}_{i+1} 
- {H\over 2} (S_{iz} + S_{i+1z}) ,\label{model}
\end{equation}
where $J$ is the spin exchange constant and $H$ is an applied
magnetic field\cite{model}.  In this paper, we consider
antiferromagnetic spin chains only.  We use units in which $J
=1$.

The spin-1/2 Heisenberg model is integrable by Bethe ansatz.
Many of its thermodynamic quantities, for example, the
specific heat and the spin susceptibility, can be calculated
by solving the Bethe ansatz equations.  The conformal field
theory is also very useful in analysing the low temperature
low field thermodynamic properties, since the S=1/2 Heisenberg 
model is equivalent to the k=1 Wess-Zumino-Witten nonlinear 
$\sigma$ model.

Higher spin Heisenberg models are not at present 
amenable to rigorous solution. The finite temperature properties of
the model were studied mainly through 
transfer matrix \cite{Betsuyaku},
quantum Monte Carlo \cite{Yamamoto93,Kim97,Kashurnikov98},
and other approximate method \cite{Narayanan,Moukouri}.

At zero field, the ground state of ${\hat H}$ is a spin
singlet with zero spin magnetization.  At finite
field, the magnetization of the ground state becomes finite
and increases with increasing $H$.  There is a critical
field $H_{c2}$ beyond which all spins are fully polarized at
zero temperature.  If we denote $E(N, S)$ as the lowest
eigenvalue of a $N$-site Heisenberg chain with total spin
$S$ at $H=0$, then it is straightforward to show that
$H_{c2} = E(N, S_{max}) - E(N, S_{max}-1) = 4S $,
independent on $N$.  Below $H_{c2}$, a canted Neel order,
namely a state which has both ferromagnetic order along the
z-axis and antiferromagnetic order in the xy plane, exists
at sufficiently low temperature, and the pitch vector (i.e.
the value of the momenta at which the static structure
factor shows a peak) decreases continuously from $\pi$ to 0
with increasing $H$.

Integer spin chains can be described by the quantum
nonlinear $\sigma$ model.  It is from the study of this
model that Haldane made the famous `Haldane conjecture'.
The ground state of the $O(3)$ $\sigma$ model has a finite
excitation gap and consequently a finite correlation length.
The application of a magnetic field causes a Zeeman
splitting of the triplet with one member crossing the ground
state at a critical field $H_{c1}$ whose value is equal to
the excitation gap $\Delta$.  When $H<H_{c1}$, the Haldane
gap persists and the ground state is still a non-degenerate
spin singlet state.  When $H_{c1}< H<H_{c2}$, the ground
state has a nonzero magnetization with gapless excitations.
Thus across $H_{c1}$, an integer spin system undergoes a
commensurate to incommensurate transition.  This is an
interesting feature which is absent in a half odd integer
spin system.  Evidence for such transitions has been used to
identify the Haldane gap in NENP\cite{nenp88} and other
quasi-1d spin compounds\cite{Granroth96}.  Just above
$H_{c1}$, the ground state can be regarded as a Bose
condensate of the low energy boson.  Varying the magnetic
field is equivalent to varying the chemical potential for
this boson and the (uniform) magnetization corresponds to the
boson number\cite{Affleck91}.

\section{TMRG} \label{sec2}

In this section, we discuss briefly the TMRG method.  A more
detailed introduction to the method can be found from Refs.
\cite{Bursill96,Wang97}.

The TMRG method starts by mapping a 1d quantum system onto a
2d classical one with the Trotter-Suzuki decomposition and 
represents the partition function as a trace of a power 
function of virtual transfer matrix $T_M$:
\begin{equation} 
Z = {\rm Tr} e^{-\beta H} =
\lim_{M\rightarrow \infty} {\rm Tr}T_M^{N/2},
\end{equation}
where $M$ is the Trotter number. $T_M$ is defined by 
an inner product of 2M local transfer matrices
\begin{eqnarray}
&&
\langle \sigma_3^1...\sigma_3^{2M}| T_M |
\sigma_1^1...\sigma_1^{2M} \rangle  
\nonumber\\
&=&
\sum_{\{\sigma_2^k\}} \prod_{k=1}^M t(\sigma_3^{2k-1}
\sigma_3^{2k} |\sigma_2^{2k-1}\sigma_2^{2k}) 
t(\sigma_2^{2k}\sigma_2^{2k+1} |\sigma_1^{2k}\sigma_1^{2k+1}),
\nonumber
\end{eqnarray}
where 
$$t(\sigma_{i+1}^k \sigma_{i+1}^{k+1} |\sigma_i^k
\sigma_i^{k+1}) = \langle -\sigma_i^{k+1},
\sigma_{i+1}^{k+1}| e^{-\tau h_i} |\sigma_i^k,
-\sigma_{i+1}^k \rangle  $$
and $\tau =\beta / M$. $|\sigma_i^k\rangle$ is an eigenstate 
of $S_i^z$ and $\sigma_i^k$ is the corresponding eigenvalue: 
$S_i^z |\sigma_i^k\rangle = \sigma_i^k |\sigma_i^k\rangle$.  
The superscripts and subscripts in $ T_M$ and
$t$ represent the spin positions in the Trotter and real
space, respectively.

$T_M$ conserves the total spin in the Trotter space, i.e.
$\sum_k\sigma_i^k$.  Thus $T_M$ is block diagonal according
to the value of $\sum_k\sigma_i^k$\cite{Wang97}.  For the
S=1/2 Heisenberg model, it was shown rigorously that the
maximum eigenstate of $T_M$ is non-degenerate and in the
$\sum_k \sigma_i^k=0$ subspace, irrespective of the sign of
J and the value of $H$ \cite{Koma89}.  When S $>$ 1/2, we found
numerically that the maximum eigenvectors of $T_M$ are also
in the $\sum_k \sigma_i^k=0$ subblock.

In the thermodynamic limit, the free energy per spin, is given by
\begin{equation}
F= - \lim_{N\rightarrow \infty} 
{1\over N\beta} \ln Z = 
-\frac 1{2\beta}\lim_{M\rightarrow \infty} \ln \lambda_{max} ,
\end{equation}  
where $\lambda_{max}$ is the maximum eigenvalue of $T_M$.
From the derivatives of $F$ one can in principle calculate
all thermodynamic quantities.  The internal energy $U$ and
the spin magnetization $M_z$ could, for example, be
calculated from the first derivative of $F$ with respect to
$H$ and $T$, respectively.  However, numerically it is
better to calculate $U$ and $M_z$ directly from the
eigenvectors of $T_M$\cite{Wang97}.  The spin susceptibility
$\chi=\partial M_z / \partial H$ and the specific heat $C =
\partial U / \partial T$ can then be calculated by numerical
derivatives.  The specific heat such determined is generally
found to be less accurate than, for example, the
susceptibility data at low $T$.  The reason for this is that
$U$ changes very slowly with $T$ (or equivalently $C$ is
very small) at low $T$, and a small error in $U$ would lead
to a relative large error in $C$.

The correlation length of the spin-spin correlation
functions, defined by $\xi_\alpha^{-1} = - {\rm lim}_{L
\rightarrow \infty } {\rm ln} \langle S_{i, \alpha}S_{i+L,
\alpha} \rangle$, can also be calculated from this method.
The longitudinal and transverse correlation lengthes are
determined by 
\begin{eqnarray}
\xi_z^{-1} & = & {1\over 2} \lim_{M\rightarrow \infty} 
{\rm ln} {\lambda_{max} \over 
\lambda_2 } , \label{c_z} \\
\xi_x^{-1} & = & {1\over 2} \lim_{M\rightarrow \infty} 
{\rm ln} {\lambda_{max} \over 
\lambda_1 }, \label{c_x}
\end{eqnarray}
where $\lambda_2$ is the second largest eigenvalue of $T_M$
in the subspace $\sum_k \sigma_k^{i} = 0$ and $\lambda_1$ is
the largest eigenvalue of $T_M$ in the subspace $\sum_k
\sigma_k^{i} = \pm 1$.

\begin{center}
\begin{figure}
\begin{picture}(160,145)
{\epsfig{file = 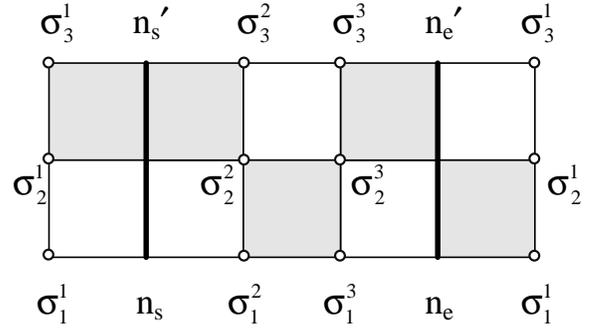, height = 2.8cm, %
                bbllx = 4.5cm, bblly=18.5cm, %
                bburx = 10.65cm, bbury=21.8cm} }
\end{picture}
\caption{A pictorial representation of a superblock, which
consists of system and environment blocks connected by three
added spins to form a periodic chain.  $n_s$ and $n_e$ are
indices for the system and environment blocks, respectively.
Initially the system block contains two spins and the
environment block contains only one spin.  At each iteration
the system and environment blocks are augmented by adding
$\sigma^1$ to each of the blocks.  The augmented system
${\tilde n}_s$ is formed by $n_e\oplus \sigma^1 $, and the
augmented environment ${\tilde n}_e$ is formed by $\sigma^1
\oplus n_s$.}
\label{supblock} 
\end{figure}
\end{center}

Figure \ref{supblock} shows the configuration of superblock
used in our calculation.  This configuration of superblock
is different from those used in Refs.
\cite{Bursill96,Wang97}.  The advantage for forming
the superblock in such a way is that the transfer matrix
$T_M$ in this case can always be factorized as a product of
two sparse matrices (which are block diagonal with respect
to $n_s$ and $\sigma_3 \otimes n_e$, respectively).  To
treat these two sparse matrices instead of $T_M$ itself
allows us to save both computer memory space and CPU time.

We compute the maximum eigenvalue, $\lambda_{max}$, and the
corresponding right and left eigenvectors, $|\psi^R\rangle$
and $\langle \psi^L|$, of $T_M$ using an implicitly
restarted Arnoldi method\cite{Sorensen92}.  This method is
more efficient than the power method which we used
before\cite{Bursill96,Wang97}.  $\lambda_{1,2}$ can also be
calculated from this method, but their truncation errors 
are generally larger than those of $\lambda_{max}$. 
Thus the correlation lengthes determined from Eqs. 
(\ref{c_z}, \ref{c_x}) are generally expected to be 
less accurate than the free energy or other thermodynamic
quantities.

In the TMRG method, the density matrix for the augmented
system or environment block is non-symmetric, which is
different than in the zero-temperature DMRG method.
Numerically it is much more difficult to treat accurately a
nonsymmetric matrix than a symmetric one because the errors
in $|\psi^R \rangle$ and $\langle \psi^L |$ may affect the
the (semi-) positiveness of the density matrix and increase
the truncation error of the TMRG.

The TMRG treats directly an infinity spin chain.  There is
therefore no finite size effect.  The error caused by the
finiteness of the Trotter number (or $\tau$) is of order
$\tau^2$, which is generally very small.  The error resulted
from the truncation of basis states is difficult to
estimate.  A rough estimation for this type of error can be
obtained from the value of truncation error, which is
smaller than $10^{-3}$ in all our calculations.  More
accurate results can be obtained simply by extrapolating the
results with respect to both $\tau$ and the number of states
retained $m$.

\section{Results} \label{sec3}

\subsection{S=1/2}

The S=1/2 Heisenberg model is by far the best understood
spin system.  In the absence of field, the ground state is
massless and the Bethe ansatz result for the ground state
energy is $(1/4-\ln 2)$.  The lowest excitatins states are
spin triplets with a spin wave spectrum \cite{Cloizeaux62}
$\varepsilon (k) = (\pi /2 ) |\sin k |$.  Above this lower
boundary of excitations, there is a two-parameter continuum
of spin wave excitations with an upper boundary given by
\cite{Yamada69} $\varepsilon (k) = \pi |\sin (k/2) |$.
There are other excitations above this upper boundary.

The specific heat of the model was first calculated
numerically by Bonner and Fisher\cite{Bonner64}.  They found
that at low temperature $C\sim 0.7T$ when $ H=0 $.  Later,
Affleck\cite{Affleck86}, using the conformal field theory,
found that $C = (2/3) T$.  The zero field zero temperature
magnetic susceptibility is $\chi = 1/ \pi^2$.  This result
was first obtained by Griffiths with the Bethe ansatz and
numerics\cite{Griffiths64}.  Recently, Eggert {\it et
al}\cite{Eggert94} found that the zero field susceptibility
approaches to its zero temperature value logarithmically
when $T< 0.01$,
\begin{equation}
\chi = {1\over 2\pi v} \left[ 1 + \left(2\ln {T_0 \over T} 
\right)^{-1} \right],
\end{equation}
where $v= (\pi /2) $ is the spin wave velocity and
$T_0\approx 7.7$.

\begin{figure}
\begin{center}
\leavevmode\epsfxsize=8.6cm
\epsfbox{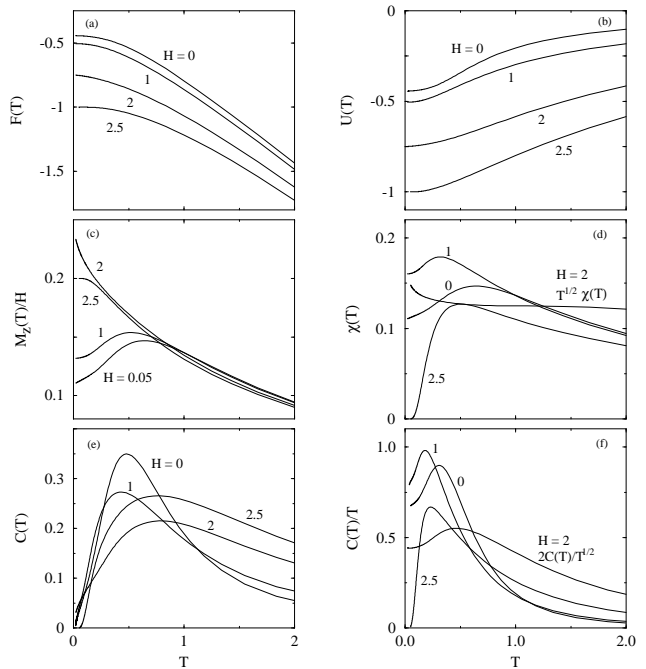}
\caption{ (a) Free energy $F(T)$, (b) internal energy
$U(T)$, (c) ratio between the magnetization $M_z(T)$ and the
field $H$, (d) spin susceptibility $\chi (T)$, (e) specific
heat $C(T)$, and (f) $C(T)/T$, per lattice site, for the
spin-1/2 Heisenberg model in four applied fields, $H$ = 0
(or 0.05), 1, 2, and 2.5.  When $H=2$, 
$T^{1/2}\chi (T)$ and $2 C(T) /T^{1/2}$, instead of $\chi
(T)$ and $C(T)/T$, are shown in (d) and (f), respectively.
$\tau =0.1$ and $m=81$ are used in the TMRG calculations.  }
\label{half}
\end{center}
\end{figure}

Figure \ref{half} shows the TMRG results for a number of
thermodynamic quantities of the S=1/2 antiferromagnetic
Heisenberg model in various magnetic fields with $m=81$ and
$\tau =0.1$.  At zero field, the TMRG reproduces accurately
the results which were previously obtained by the Bethe
ansatz or conformal field theory.  The extrapolated zero
field zero temperature values of the internal energy $U(T)$
(i.e.  ground state energy), the spin susceptibility $\chi
(T)$ and the linear coefficient of the specific heat $C(T)$,
are -0.443, 0.109 and 0.66, respectively.  In all the fields
which we studied, the peak position of $\chi (T)$ is located
at a temperature which is about twice of the peak
temperature of $C(T) / T$.  This is due to the fact that
$\chi (T)$ is a measure of two-particle excitations and
$C(T)/T$ is only a measure of one-particle density of
states.  The maximum of $\chi$ when $H=0$ is approximately
equal to $0.147$ at $T=0.64$, consistent with the Bethe
ansatz calculation\cite{Eggert94}.

There is a significant change in the temperature dependences
of $\chi$ and $C$ when $H$ is below and above a critical
field $H_{c2} = 2$.  Below $H_{c2}$, both $C/T$ and $\chi$ are
finite at zero temperature, which shows that gapless spin
excitations with a finite low energy density of states exist
in this regime.  When $H=H_{c2}$, both $\chi$ and $C/T$ vary as
$ 1/ \sqrt{T} $ at low temperature; the extrapolated value
of $\sqrt{T}\chi$ and $C/ \sqrt{T}$ at zero temperature are
$0.152$ and $0.22$, respectively.  The divergency of $\chi$
and $C / T$ at $T=0$ implies that the density of states of
spin excitations is divergent at zero energy when $H = H_{c2}$.
Above $H_{c2}$, there is a gap in the excitation spectrum as
both $\chi$ and $C/T$ drop to zero exponentially at low
temperatures.  The value of the gap estimated from the low
temperature behavior of $\chi$ and $C$ grows linearly with
$H-H_{c2}$.

\begin{figure}
\begin{center}
\leavevmode\epsfxsize=7cm
\epsfbox{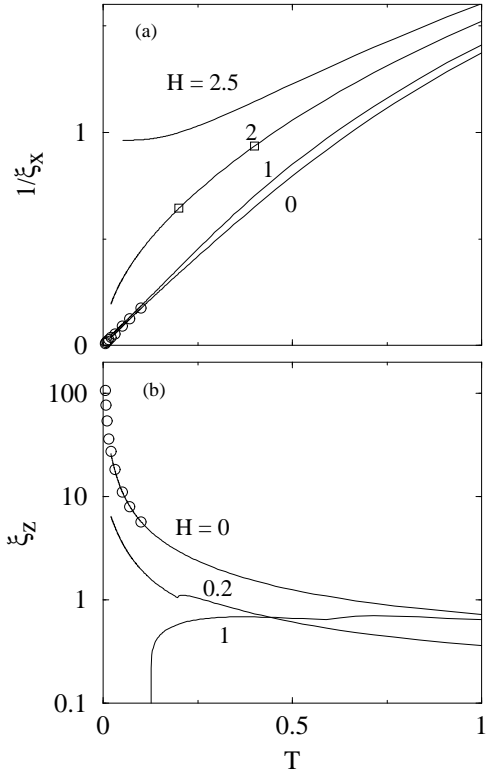}
\caption {(a) $1/ \xi_x$ and (b) $\xi_z$ vs $T$ for the 
spin-1/2 Heisenberg models.  $\tau =0.1$ 
and $m=81$ are used in the TMRG calculations.  The circles
and squares 
are thermal Bethe ansatz results.  }
\label{corrhalf}
\end{center}
\end{figure}

Figure (\ref{corrhalf}) shows the longitudinal and
transverse correlation lengthes of the S=1/2 model in
different applied fields.  When $H=0$, $\xi_x=\xi_z$
diverges at zero temperature and the slope of $\xi_z^{-1}$
at low temperature is approximately equal to 2, in agreement
with the thermal Bethe ansatz as well as the k=1 WZW
$\sigma$ model result\cite{Nomura91}
\begin{equation}
\xi_z^{-1} = T\left[ 2 - \left( \ln {T_0\over T}\right)^{-1}
\right]
\end{equation}

In the presence of magnetic field, $\xi_z$ is substantially
suppressed and becomes finite at zero temperature.  As the
correlation length is inversely proportional to the energy
gap of excitations, the finiteness of $\xi_z$ at $T=0$ means
that the longitudinal spin excitation modes are massive in a
field.  There is a small dip in the curve of $\xi_z$ at
$T\sim 0.4$ when $H = 0.2$ or at $T\sim 0.6$ when $H=1$.
This dip feature of $\xi_z$, as will be shown later, appears
also in large $S$ systems.

The effect of the applied field on the transverse spin
excitation modes is weaker than the longitudinal modes.  The
transverse spin excitations become massive only when $H >
H_{c2}$.  Below $H_{c2}$, $\xi_x$ drops to zero linearly with $T$,
as for the case $H=0$.  When $H = H_{c2}$, $\xi_x$ drops to
zero as $\sqrt{T}$.  Clearly the thermodynamics of the
Heisenberg model in a field is mainly determined by the
transverse excitation modes at low temperatures.

A simple understanding of the above results can be obtained
from an equivalent spinless fermion model of the S=1/2
Heisenberg model:  
\begin{eqnarray} {\hat H}& =&
\sum_i\left[- {1\over 2} (c_{i+1}^\dagger c_i +c_i^\dagger
c_{i+1}) + (H-1)c_i^\dagger c_i \right.  \nonumber \\ & &
\left.  + ({1\over 4} - H) + c_i^\dagger c_i
c_{i+1}^\dagger c_{i+1}\right] , \label{fermion}
\end{eqnarray} 
where $c_i$ a spinless fermion operator which
is linked to the $S=1/2$ spin operator by the Jordan-Wigner
transformation $c_i = S_i^+ \exp ( i \pi \sum_{l < i}
c_l^\dagger c_l ) $.  The magnetic field is equivalent to a
chemical potential for the fermions.  When H=0, the ground
state has zero uniform magnetization, corresponding to a
half filled fermion band.  As $H$ increases, the ground
state is ferromagnetically polarized and the Fermi energy
shifts down.  When $H$ is smaller than the critical field
$H_{c2}$, the ground state of this fermion model has no gap
but the spin orientation is canted, i.e.  in an
incommensurate state.  The pitch angle of this
incommensurate ground state, namely the wave vector at which
the maximum of the spin-spin correlation function appears,
can be estimated from the Fermi momentum of non-interacting
fermions as $2 k_F = [1 - 2 M_z(H)] \pi$.  This value of
$k_F$ is not normalized by interactions according to the
Luttinger theorem.

Above $ H_{c2}$, there are no fermions at the ground state.
At low temperature, the number of fermions excited from the
ground state are rare.  Thus, as a good approximation, the
interaction term in (\ref{fermion}) can be ignored at low
temperatures.  For the non-interacting system, the energy
dispersion of fermion excitations from the ground state is
given by $\varepsilon_k = H -  (1+ \cos k)$, which has a
gap of $H-H_{c2}$.  When $H = H_{c2}$, $\varepsilon_k =  (1
- \cos k)$, the density of states of excited fermions varies
as $ 1/ \sqrt{ \varepsilon }$ at low energy.  From the
standard theory of noninteracting fermions, it is
straightforward to show that this singular density of states
will cause both $\chi$ and $C/T$ to diverge as $1/ \sqrt{T}$
at low temperature.  When $H > H_{c2}$, there is a gap in
the fermion excitations, both $\chi $ and $C$ should
decrease exponentially at low temperature.  These results
are just what we found in Figure \ref{half}.

\subsection{S=1} 

\begin{figure}
\begin{center}
\leavevmode\epsfxsize=8.6cm
\epsfbox{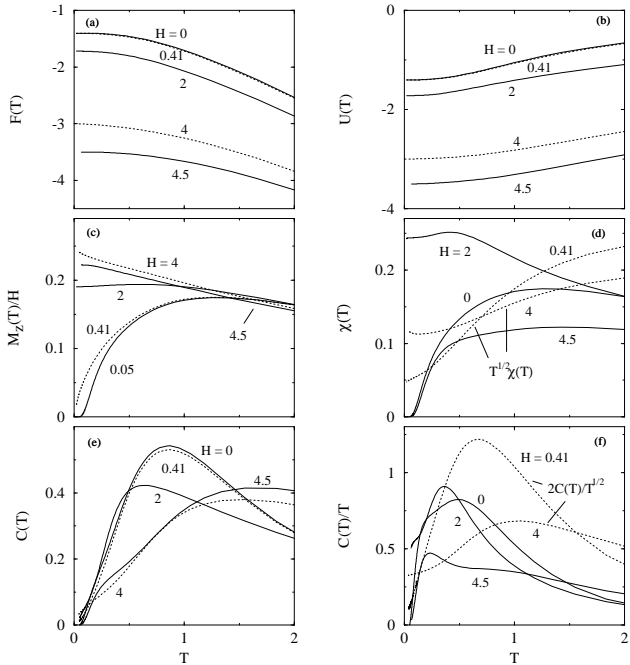}
\caption{ Thermodynamic quantities for the spin-1 Heisenberg
model in five applied fields, $H$ = 0 (or 0.05), 0.4105, 2,
4, and 4.5.  When $H=0$ and 0.4105, $F(T)$ and $U(T)$ are
almost indistinguishable in the figure.  When $H=0.4105$ or
4, $T^{1/2}\chi (T)$ and $2 C(T) /T^{1/2}$ (dotted lines)
are shown in (d) and (f), respectively.  $\tau =0.1$ and
$m=100$ are used in the TMRG calculations.  }
\label{one}
\end{center}
\end{figure}

As mentioned previously, 
there are two critical fields in the spin-1 Heisenberg
model, $H_{c1}$ and $H_{c2}$:  below $H_{c1}$, the ground
state is a massive spin singlet; above $H_{c2}$, the ground
state becomes a fully polarized ferromagnetic state; between
$H_{c1}$ and $H_{c2}$, the ground state is massless and has
a finite magnetization.  The temperature dependence of
thermodynamic quantities of the S=1 model below, above and
(approximately) at these critical fields is shown in Fig.
\ref{one}.

At zero field both $\chi (T)$ and $C(T)$ drop exponentially
with decreasing $T$ at low temperatures duo to the opening
of the Haldane gap.  In this case the ground state energy
extrapolated from the internal energy $U(T)$ is -1.4015, in
agreement with the zero-temperature DMRG result
\cite{White93}.  The ground state excitation gap $\Delta$
can be determined from the temperature dependence of $\chi$
and $C$ at low temperatures.  If we adapt the ansatz that
the low-lying excitation spectrum has approximately the form
\cite{Sorensen93}
\begin{equation}
\varepsilon (k) = \Delta + {v^2 \over 2\Delta} (k-\pi )^2 +O(|k-\pi|^3),
\end{equation}
where $v$ is the spin wave velocity, it is then 
straightforward to show that, when $T \ll \Delta$, the spin
susceptibility and the specific heat are
\begin{eqnarray}
\chi (T) & \approx & {1\over v} \sqrt{2\Delta \over \pi T} 
e^{-\Delta / T}, \label{sus1}
\\
C(T) & \approx &
{3 \Delta  \over v \sqrt{ 2 \pi } } \left( 
{\Delta  \over T }\right)^{3/2} e^{-\Delta  / T},  
\end{eqnarray}
irrespective of the statistics of the excitations. Taking the 
ratio between $\chi (T)$ and $C(T)$ gives
\begin{equation}
\Delta = \lim_{T\rightarrow 0} \sqrt{2TC(T)\over 3\chi (T)}.
\label{gapeq}
\end{equation}
This is a very useful equation for determining $\Delta$, 
especially from the point of view of experiments, since both
$\chi (T)$ and $C (T)$ are experimentally measurable quantities.

\begin{figure} 
\begin{center} 
\leavevmode\epsfxsize=8cm
\epsfbox{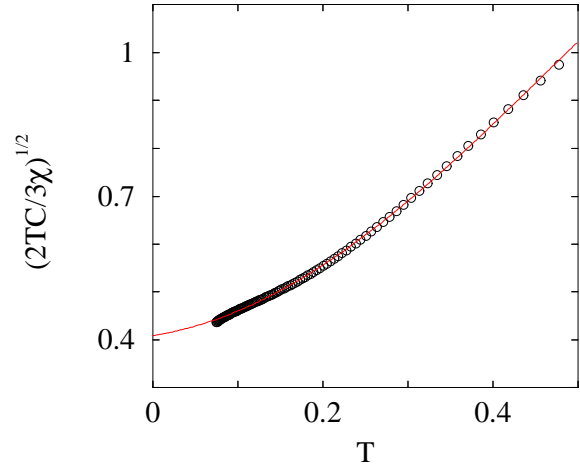}  
\caption {$({2TC(T)/3\chi (T)})^{1/2}$ vs $T$ for the 
S=1 Heisenberg model. ($m=100$ and $\tau =0.1$.)}
\label{gap} 
\end{center} 
\end{figure}

Fig.  \ref{gap} shows $\sqrt{2TC(T)/ 3\chi (T)}$ as a
function of $T$ for the S=1 Heisenberg model at zero field.
By extrapolation, we find that $\Delta \sim 0.41$ in
agreement with the zero temperature DMRG \cite{White92} and
exact diagonalization \cite{Golinelli94} results.  Given
$\Delta$, the value of $v$ can be found from Eq.
(\ref{sus1}) in the limit $T\rightarrow 0$.  The value of
$v$ we found is $\sim 2.45$, consistent with other numerical
calculations \cite{Sorensen93}.

At the two critical fields, $H_{c1}\sim 0.4105$ and
$H_{c2}=4$, both $\chi (T)$ and $C(T)/T$ diverge as
$T^{-1/2}$ at low temperatures.  This divergence, as for the
S=1/2 case, is due to the square-root divergence of the
density of states of the low-lying excitations.  At
$H_{c1}$, one branch of the S=1 excitations becomes massless
and the low-energy excitation spectrum is approximately
given by
\begin{equation}
\varepsilon (k) \sim {v^2\over 2\Delta} (k-\pi)^2.
\end{equation}
If we assume that these excitations are fermion-like, i.e.
satisfy the Fermi statistics (a short-range interacting Bose
system is equivalent to a system of free fermions), then it
is simply to show that when $T\ll \Delta$
\begin{eqnarray}
\chi (T) & \approx & { 1 \over 3\pi v }  \sqrt{2\Delta \over T},
\\
C (T) & \approx & {1 \over 2\pi v} \sqrt{2\Delta T} .
\end{eqnarray}
Thus in the limit $T\rightarrow 0$, 
\begin{eqnarray}
T^{1/2}\chi (T) & = & { \sqrt{2\Delta} \over 3\pi v} 
\sim 0.4 , \label{susT0} \\
T^{-1/2}C (T) & = & {\sqrt{2\Delta } \over 2\pi v} \sim 0.6.
\label{heatT0} 
\end{eqnarray}

By extrapolation, the TMRG result gives $T^{1/2}\chi
|_{T\rightarrow 0} \sim 0.45$, which is close to that given by Eq.
(\ref{susT0}).  The value of $T^{-1/2} C|_{T\rightarrow 0}$
is difficult to determine accurately from the TMRG result
since the error of $C(T)$ at low temperature is larger than
$C(T)$ itself.

\begin{figure} 
\begin{center} 
\leavevmode\epsfxsize=7cm
\epsfbox{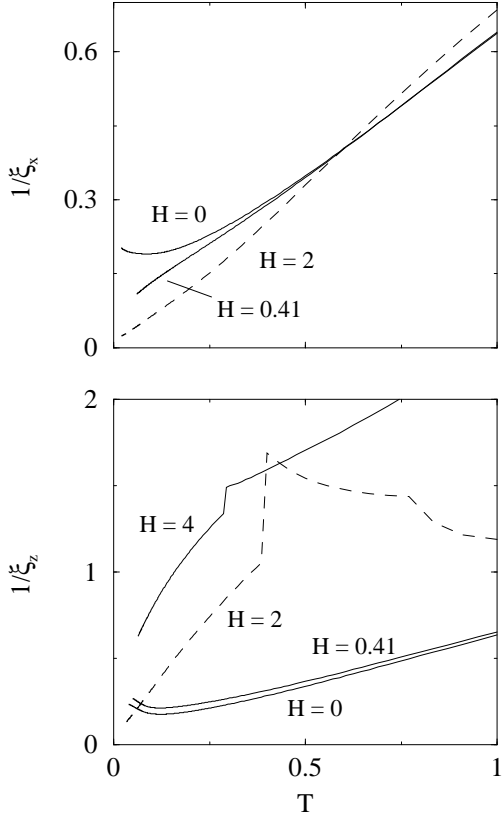} 
 \caption {(a) $\xi_x^{-1}$ and (b) $\xi_z^{-1}$ vs $T$ for
the spin-1 Heisenberg models.  $\xi_x^{-1}(H=0)$ and
$\xi_x^{-1}(H=0.4105)$ are almost indistinguishable when $T
> 0.6$.  $\tau =0.1$ and $m=100$ are used in the TMRG
calculations.  }
\label{corrone} 
\end{center} 
\end{figure}

Fig.  \ref{corrone} shows $\xi^{-1}_x$ and $\xi_z^{-1}$ of
the S=1 model in different applied fields.  Because of the
Haldane gap, the longitudinal correlation
length is finite even at zero field.  Thus the longitudinal
excitation mode is always gapful.  Above $H_{c1}$ but below
$H_{c2}$, the transverse correlation length
diverges at zero temperature.  Thus the transverse mode is
massless in this field regime.  We note, however, that
$\xi_{x,z}^{-1}$ at $H=0$ (similarly $\xi_z^{-1}$ at
$H=0.4105$) does not decrease monotonically at low
temperatures, actually it starts to raise
 below a temperature $T^*$.  This non-monotonic behavior
of $\xi^{-1}$ at $H=0$ seems only due to the truncation error.  
Our preliminary calculation shows that $T^*$
decreases with increasing $m$.  But by far we still do not
know if $T^*$ is zero in the limit $m\rightarrow \infty$. 
Further investigation to the low temperature behavior of 
$\xi$ is needed.  If we do an extrapolation using 
the TMRG data of $\xi^{-1}_x$ above $T^*$, we find that
$\xi_x (T=0) \sim 6.0$, which is consistent with the zero
temperature DMRG result \cite{White93}. If, however, the 
TMRG data below $T^*$ are also included in the extrapolation, 
we find that $\xi_x (T=0)$ is only $\sim 4.4$.

When $H > H_{c1}$, $\xi_{z}^{-1}$ drops rather sharply at
some temperatures.  These sharp drops of $\xi^{-1}_z$ happen
when the second and third eigenvalues of the transfer matrix
with $\sum_k\sigma^k = 0$ cross each other.  The physical
consequence of these sudden changes in the longitudinal
correlation length is still unknown.

\subsection{S=3/2}

\begin{figure}
\begin{center}
\leavevmode\epsfxsize=8.6cm
\epsfbox{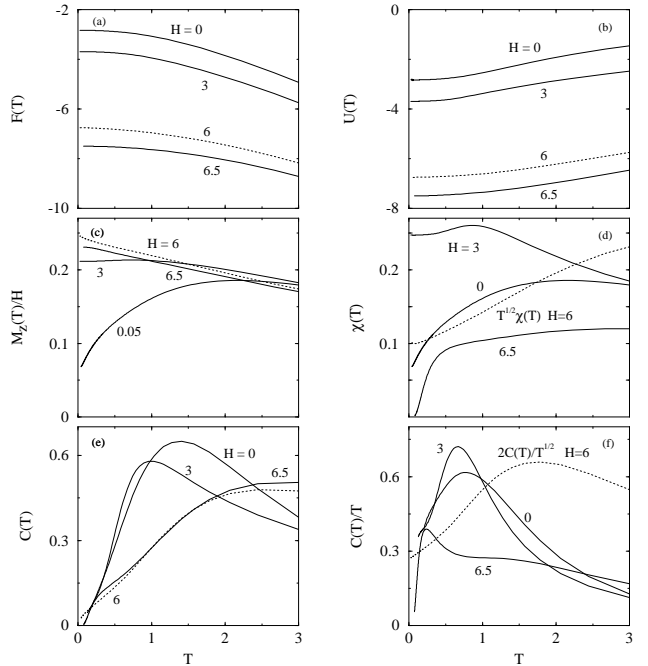}
\caption{ Thermodynamic quantities for the spin-3/2 Heisenberg
model in four applied fields, $H$ = 0 (or 0.05), 3,
6, and 6.5.  When $H=6$, 
$T^{1/2}\chi (T)$ and $2 C(T) /T^{1/2}$ (dotted lines) 
are shown in (d) and
(f), respectively.  $\tau =0.075$ and $m=81$ are used in the
TMRG calculations.  }
\label{half3}
\end{center}
\end{figure}

The thermodynamic behaviors of the S=3/2 Heisenberg model,
as shown in Fig.  \ref{half3}, are similar to those of the
S=1/2 model.  When $H< H_{c2}=6$, $\chi$ is always finite at
zero temperature, indicating that the ground state is
massless; above $H_{c2}$, the ground state is fully
ferromagnetic polarized and a gap is open in the excitation
spectrum; at $H_{c2}$, both $\chi (T)$ and $C/T$ diverge as
$T^{-1/2}$.  At zero field,  the susceptibility data, extrapolated 
to zero temperature, gives $\chi (0)\sim 0.67$, consistent 
with recent numerical calculations \cite{Kim97,Moukouri}.

\begin{figure}
\begin{center}
\leavevmode\epsfxsize=8cm
\epsfbox{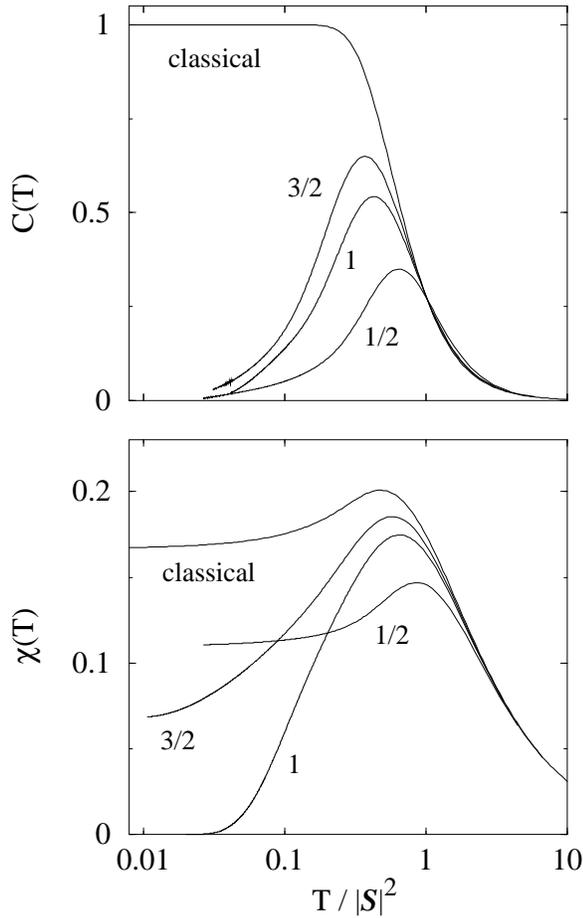}
\caption {The zero field specific heat and susceptibility vs
$T/|{\bf S}|^2$ for the S=1/2, 1, and 3/2 spin chains
($|{\bf S}|^2 = {\rm S(S+1)}$).  The corresponding results
for the classical Heisenberg model ($|{\bf S}|^2=1$) are
also shown for comparison.  }
\label{zerofield}
\end{center}
\end{figure}

The crossover from quantum to classical behavior can be
clearly seen (Fig.  \ref{zerofield}) by comparing the zero
field susceptibility and specific heat of the S=1/2, 1 and
3/2 spin chains with the corresponding results of the
classical Heisenberg spin chain:  \cite{Fisher64}
\begin{eqnarray}
\chi (T) &=& {1\over 3T} {1-u(T)\over 1+u(T)}, 
\quad u(T)=\coth {1\over T} - T ,\\
C(T) &= & 1 - {1\over T^2 \sinh^2 (1/T)} .
\end{eqnarray}
At high temperatures, $T/S(S+1) > 1$, the quantum results
approach asymptotically to the classical ones.  The
agreement between the quantum and classical results persist
down to progressively lower temperatures as S increases.  At
low temperatures, however, the difference between the
results of the S=3/2 system and those of the classical model
is still very large, indicating the importance of the
quantum effects in the study of quantum spin chains.  (Note
for the classical Heisenberg model, $C(T)$ does not vanish
at zero temperature.  This is a unrealistic feature of this
model.)

\subsection{Zero temperature magnetization}

\begin{figure}
\begin{center}
\leavevmode\epsfxsize=8cm
\epsfbox{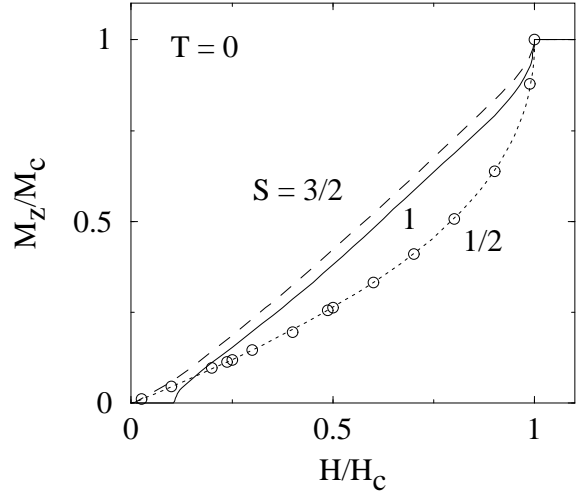}
\caption {Normalized zero temperature magnetization $M_z$ as
a function of $H / H_{c2}$.  The TMRG results for S=1/2
(circles), S=1 (solid line), and S=3/2 (dashed line) are
obtained by extrapolation from the low $T$ values of $M_z$
with $(m, \tau) = (81, 0.1)$, (81, 0.1), and (60, 0.75),
respectively.  Dotted line is the Bethe ansatz result for
the S=1/2 system.
}
\label{MT0}
\end{center}
\end{figure}

Figure \ref{MT0} shows $M_z (T=0)$, extrapolated from the
finite temperature TMRG data of $M_z(T)$, for the S=1/2, 1,
and 3/2 systems.  For comparison the Bethe ansatz
result\cite{Griffiths64} for the S=1/2 Heisenberg model is
also shown in the figure. With increasing S, 
we found that $M_z$ tends to approach to its classical
limit ($S \rightarrow \infty$) where $M_z$ increases
linearly with $H$, i.e.  $M_z / S = H / H_{c2}$.

For the S=1/2 system, the TMRG result agrees well with the
Bethe ansatz one.  A least square fit to the curve of $M_z$
up to the fourth order term of $\sqrt{H_{c2} -H}$ gives $M =
M_c - a_1 \sqrt{ H_{c2}-H} + a_2 (H_{c2}-H) -
a_3(H_{c2}-H)^{3/2} + a_4 (H_{c2}-H)^2$, with $a_1\approx
0.448$, $a_2 \approx 0.123$, $a_3 = 0.05$ and $a_4=0.00744$.
$a_1$ agrees very accurately with the exact value $\sqrt{2}/
\pi$.  In the weak field limit, the asymptotic behavior of
$M_z$ is 
\begin{equation}
 M_z\approx {H \over \pi^2} \left( 1 -{1 \over 2 \ln
(H/H_{c2}) } \right),
\end{equation} 
as predicted by the Bethe Ansatz theory \cite{Yang66,Babujian83}.

For the S=1 model, $M_z(T=0)$ becomes finite when
$H>H_{c1}$.  In a very narrow regime of field near $H_{c1}$,
$M_z (0)$ varies approximately as $\sqrt{H - H_{c1}}$, in
agreement with the prediction of the Bose condensation
theory \cite{Affleck91}.  But the difference between the 
result of the Bose condensation theory 
\begin{equation}
M_z (T=0) \approx {\sqrt{2 (H - H_{c1}) \Delta }\over \pi v} 
\end{equation} 
and that of the TMRG becomes already significant when 
$H-H_{c1} = 0.04$.

Near $H_{c2}$, $M_z$ approaches to its saturation value
$M_c=S$ as a function of $\sqrt{H_{c2} - H}$ for the three
spin systems we study.  For the S=1/2 and 1 systems, the
TMRG results agree very accurately with the Bethe ansatz
result \cite{Yang66,Parkinson85}
\begin{equation}
{M_z \over S} = 1 -{2 \over \pi S} \sqrt{ 1 - {H \over H_{c2}}}.
\label{MzT=0} 
\end{equation}

For the S=3/2 system, the asymptotic regime of $H$ is 
very narrow, we cannot do a detailed comparison between 
the TMRG result and Eq. (\ref{MzT=0}). The magnetization 
curve does not show a plateaus at $M_z = 1/2$, 
in agreement with other studies \cite{Oshikawa97}.

\subsection{Staggered susceptibility}

To calculate the staggered susceptibility, we add a
staggered magnetic field $H_s$ to the Hamiltonian ${\hat
H}$.  The staggered magnetization is then calculated in a
way similarly to the calculation of the uniform
magnetization.  The staggered susceptibility $\chi_s$ is
obtained by differentiating the staggered magnetization with
respect to $H_s$.  For half-odd-integer spin chains $\chi_s$
diverges as $T^{-1}$ at low temperatures.  Thus the
staggered magnetization becomes saturated at low temperatures
when $\chi_s (T) H_s > S$, since the maximum value of the
staggered magnetization per spin is S.  This means that to
estimate accurately the zero field staggered susceptibility
the staggered field used should satisfy the condition $H_s
\ll S/\chi_s(T_{min})$, where $T_{min}$ is the lowest
temperature to study.

\begin{figure}
\begin{center}
\leavevmode\epsfxsize=8cm
\epsfbox{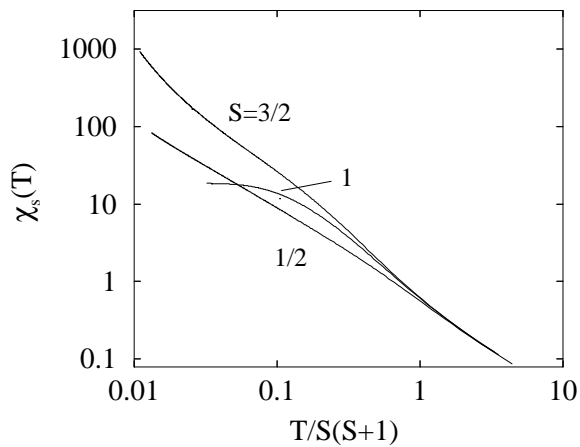}
\caption {the staggered 
susceptibility $\chi_s (T)$ at zero field. The parameters
used for the S=1/2, 1, and 3/2 chains are ($H_s$, 
$\tau$, m) = (0.0001, 0.1, 140), (0.0001, 0.05, 100), and 
(0.0001, 0.025, 81), respectively. }
\label{stagger}
\end{center}
\end{figure}

Fig.  \ref{stagger} shows the temperature dependence of
$\chi_s$ for the S=1/2, 1, and 3/2 spin chains.  At low
temperatures, $\chi_s$ for the S=1/2 model is expected to
have the following asymptotic form
\begin{equation}
\chi_s = {D_\chi \over T} \sqrt{ \ln {T_\chi \over T} } \label{stag}
\end{equation}
according to a scaling argument \cite{Starykh97}.  
Our TMRG result agrees well with this equation at low
temperatures.  By plotting $(T \chi_s)^2$ against $\ln T$,
we find that $D_\chi \approx 0.30$ and $T_\chi \approx 10.5$, 
in agreement with a Monte Carlo calculation \cite{Kim97}.

For the S=1 model, $\chi_s$ is finite at zero temperature. 
The extrapolated zero temperature value of $\chi_s$ for the 
S=1 model is 18.55, in agreement with other numerical 
calculations \cite{Kim97,Sakai90}.

\section{Conclusion} \label{sec4}

In conclusion, the temperature dependence of the
susceptibility, specific heat, and several other quantities
of the quantum Heisenberg spin chains with spin ranging from
1/2 to 3/2 in a finite or zero applied magnetic field are
studied using the TMRG method.  At high temperatures, the
quantum results for the specific heat as well as other
thermodynamic quantities approach asymptotically to the
classical ones for both the integer and half-integer spin
systems.  At low temperatures, however, the quantum effect
is strong and the integer spin chains behave very
differently than the half-integer spin chains.  For the S=1
model, both $\chi$ and $C$ decay exponentially at low
temperatures due to the opening the Haldane gap.  For the
S=1/2 and 3/2 spin chains, there is no gap in the excitation
spectrum and both $\chi$ and $C/T$ are finite at zero
temperature.  The thermodynamics of the Heisenberg spin
chains in an applied field is mainly determined by the
transverse excitation modes.  At low temperatures, $\chi$,
$C/T$, and the transverse correlation length $\xi_x$ diverge
as $T^{-1/2}$ at $H_{c2}$ for the S=1/2 and 3/2 models and
at both $H_{c1}$ and $H_{c2}$ for the S=1 model.  This
square-root divergence indicates that the low energy spin
excitations have a square-root divergent density of states
at these critical fields.  Our data agree well with the
Bethe ansatz, quantum Monte Carlo, and other analytic or
numerical results.

\section*{Acknowledgments}

I would like to thank R.  J.  Bursill, G. A. Gehring, 
S.  J.  Qin and X. Wang for stimulating discussions.


\end{document}